\documentclass[aps,prd,showpacs,nofootinbib,eqsecnum,amsmath,amssymb,twocolumn]{revtex4-1}
\usepackage{epsf}
\usepackage{color}
\usepackage{dcolumn}% Align table columns on decimal point
\usepackage{bm}% bold math
\everymath{\displaystyle}

\begin{document} %%%%%%%%%%%%%%%%%%%%%%%%%%%%%%%%%%%%%%%%%%%%%%%%%

\title{Spin-torsion coupling and gravitational moments of Dirac fermions:
theory and experimental bounds}

\author{\firstname{Yuri N.}~\surname{Obukhov}}
\email{obukhov@ibrae.ac.ru}
\affiliation{Theoretical Physics Laboratory, Nuclear Safety Institute,
Russian Academy of Sciences, B. Tulskaya 52, 115191 Moscow, Russia}

\author{\firstname{Alexander J.}~\surname{Silenko}}
\email{alsilenko@mail.ru} \affiliation{Bogoliubov Laboratory of Theoretical Physics, Joint Institute for Nuclear Research,Dubna 141980, Russia,\\
Research Institute for Nuclear Problems, Belarusian State University, Minsk 220030, Belarus}

\author{\firstname{Oleg V.}~\surname{Teryaev}}
\email{teryaev@theor.jinr.ru} \affiliation{Bogoliubov Laboratory
of Theoretical Physics, Joint Institute for Nuclear Research,
Dubna 141980, Russia,\\ 
National Research Nuclear University ``MEPhI'' (Moscow Engineering
Physics Institute), Kashirskoe Shosse 31, 115409 Moscow, Russia}

%\date{file ``ostor12.tex", \today}

\begin {abstract}
We discuss the quantum dynamics of a Dirac fermion particle in the Poincar\'e gauge gravitational field. The minimal as well as the Pauli-type nonminimal coupling of a fermion with external fields is studied, bringing into consideration the notions of the translational and the Lorentz gravitational moments. The anomalous gravitomagnetic and gravitoelectric moments are ruled out on the basis of the covariance arguments. We derive the general Foldy-Wouthuysen transformation for an arbitrary configuration of the Poincar\'e gauge gravitational field without assuming it weak. Making use of the Foldy-Wouthuysen Hamiltonian for the Dirac particle coupled to magnetic field in a noninertial reference system, we analyze the recent experimental data and obtain bounds on the spacetime torsion.
\end{abstract}

\pacs{04.20.Cv; 04.62.+v; 03.65.Sq} \maketitle

%%%%%%%%%%%%%%%%%%%%%%%%%%%%%%%%%%%%%%%%%%%%%%%%%
\section{Introduction}
%%%%%%%%%%%%%%%%%%%%%%%%%%%%%%%%%%%%%%%%%%%%%%%%%

Poincar\'e gauge theory of gravity is a natural extension of Einstein's general relativity (GR) theory based on the gauge-theoretic ideas; see a comprehensive review in \cite{rmp,MAG,Blag,reader}. The geometrization of the gravitational physics using the principles of covariance and equivalence is similar to the geometrization of the three physical interactions (electromagnetic, weak and strong) using the Yang-Mills type of approach. There is a difference though in that the Standard Model deals with the fundamental symmetry groups acting in the internal spaces, whereas the gravity has to do with the symmetry of the spacetime.

The group of the local spacetime translations (diffeomorphisms) plays the central role in GR. This is manifest in the well known fact \cite{Feyn} that the gravitational field couples to the corresponding translational Noether current -- the energy-momentum tensor. On the other hand, the high energy physics is based on the Poincar\'e group which is a semidirect product of the translation group times the Lorentz group. The fundamental particles are classified by mass and spin which arise in the representation theory of the Poincar\'e group. The Noether theorem gives rise to the two currents, in accordance with the semidirect structure of the Poincar\'e group: the energy-momentum tensor (translational current) and the tensor of spin (rotational current). In the gauge theories of the Yang-Mills type, the principle of the local symmetry relates the existence of the gauge fields to the corresponding Noether currents. In the gauge-theoretic framework, there exists a natural extension of GR based on the Poincar\'e group, with the energy-momentum and spin currents as the sources of the gravitational field \cite{Sciama,Kibble,HonnefH,Tra}. The spacetime geometry is then characterized by a nontrivial torsion which is coupled to spin current, along with the metric coupled to the energy-momentum current.

Theory of gravity with torsion has a long history going back to 1922 when \'Elie Cartan came up with the first gravitational model \cite{Cartan}. Later it attracted much of attention in attempts to construct the unified field theories (with the notable efforts of Weyl, Einstein, Eddington, and Schr\"odinger among others \cite{Goenner}). Another important step was the development of physical models of elastic media with microstructure by Cosserats, Kroener \emph{et al} \cite{essay}. The modern understanding of the torsion and of its relation to the gravitational physics was achieved in the framework of the Poincar\'e gauge theory \cite{Sciama,Kibble,HonnefH,MAG,Blag,reader}. The Einstein-Cartan gravitational theory \cite{Tra,rmp} is the closest viable extension of GR. It is consistent with experiments on the macroscopic scales and, in particular, with all classical gravity tests within the Solar System. A possible deviation from the GR due to the contact spin-torsion interaction is only expected at extremely high densities during the early stages of universe's evolution or on the microscopic scales in the high energy particle experiments.

The post-Riemannian geometry of spacetime can be probed with the help of detectors built of the matter with microstructure. The classical point particles with spin and spinning continuous media (fluids) were extensively studied in this context \cite{WR,halb,Ritis,Kopc,OK}. The analysis of the equations of motion of extended bodies \cite{YS,PO2007} has shown that the torsion can be measured only when the matter possesses intrinsic spin. Mechanically rotating gyroscopes do not feel the torsion when matter couples minimally to the gravitational field; there is however a loophole for the nonminimal coupling case \cite{pla2013}. In the efforts to detect the spacetime torsion, the polarized material bodies and media are systematically used in the recent experiments \cite{Hunter,Lehnert}.

On the cosmological scales, the modern observations can be used to place limits on the possible torsion effects \cite{Ogino} which may qualitatively modify the early stage of universe's evolution \cite{AT,pop,kib}; for an overview of cosmology with spin and torsion see \cite{DPast}.

The general discussion of the spin-torsion classical and quantum effects can be found in Ref. \cite{Shapiro}; see Ref. \cite{Ni2010} for a more recent review.

In the present paper we consider the quantum dynamics of a Dirac fermion particle, taking into account possible spin-torsion coupling in the framework of the Poincar\'e gravity. Earlier this problem was analyzed for special gravitational field configurations in the semiclassical approximation \cite{Rumpf,HS1,Aud,Bagrov1,Bagrov2,Nomura,ryder}. Here we generalize our previous results \cite{Ob1,Ob2,PRD,Warszawa,OST,OSTRONG,ostgrav} obtained for the dynamics of fermion spin in an arbitrary torsionless gravitational field.

The structure of the paper is as follows. In Sec.~\ref{Poincare} we recall the basic facts about the gauge-theoretic approach to gravity and describe in full detail the coupling of a fermion Dirac particle to the electromagnetic and the Poincar\'e gauge gravitational field. The Foldy-Wouthuysen transformation is performed in Sec.~\ref{Hamiltonian} for an arbitrary spacetime geometry with the curvature and torsion. The possible nonminimal coupling of the Dirac particle to the Poincar\'e gauge field is discussed in Sec.~\ref{DHamiltonian}, where we demonstrate the importance of the Gordon decomposition of the Noether currents. We then specialize in Sec.~\ref{MFRF} to the dynamics of a Dirac fermion particle in the magnetic field in a rotating reference frame. In Sec.~\ref{NB} we use the theoretical findings to obtain the new bounds on the spacetime torsion from the experimental data. Finally, we summarize our conclusions in Sec.~\ref{Conclusion}. In Appendix A we describe our notations and conventions.

%%%%%%%%%%%%%%%%%%%%%%%%%%%%%%%%%%%%%%%%%%%%%%%%%
\section{Spin-torsion coupling in Poincar\'e gauge gravity}\label{Poincare}
%%%%%%%%%%%%%%%%%%%%%%%%%%%%%%%%%%%%%%%%%%%%%%%%%

Let us give a brief summary of the corresponding gauge-theoretic formalism, without going into the subtleties of constructing the gauge theory for the Poincar\'e group (technical details can be found in Refs. \cite{Sciama,Kibble,HonnefH,MAG,Blag,reader}).

At first, we recall the essential points of the Yang-Mills theory. Let $\Phi^A$ denote the matter field, and $G$ is the $N$-parameter symmetry group. Under its action, the field transforms covariantly
\begin{equation}
\delta\Phi^A = \varepsilon^I(\rho_I)^A_B\Phi^B.\label{dPhi}
\end{equation}
Here $(\rho_I)^A_B$, are the generators in the corresponding representation of $G$ (with {\scriptsize{$I,J,K$}} $= 1,\dots, N$; the range of the indices {\scriptsize{$A,B,C,\dots$}} is not important). When the infinitesimal parameters $\varepsilon^I$ are constant, the derivatives transform covariantly $\delta\,\partial_i\Phi^A = \varepsilon^I(\rho_I)^A_B\partial_i\Phi^B$. However, for the {\it local symmetry} with $\varepsilon^I=\varepsilon^I(x)$ one needs the gauge field $A^I_i$ to define
\begin{equation}
D_i\Phi^A = \partial_i\Phi^A - A^I_i(\rho_I)^A_B\Phi^B.\label{DPhi}
\end{equation}
This {\it covariant derivative} transforms homogeneously, $\delta\,D_i\Phi^A = \varepsilon^I(\rho_I)^A_BD_i\Phi^B$, provided the gauge field potential changes $\delta A^I_i = \partial_i\varepsilon^I + f^I{}_{JK}\varepsilon^JA^K_i$ under group's action. The structure constants $f^I{}_{JK}$ determine the Lie algebra of the gauge group $G$, so that the generator commutator reads $(\rho_J)^A_C(\rho_K)^C_B - (\rho_K)^A_C(\rho_J)^C_B = f^I{}_{JK}(\rho_I)^A_B$.

The gauge potential gives rise to the gauge field strength tensor
\begin{equation}
F_{ij}{}^I = \partial_iA^I_j - \partial_jA^I_i - f^I{}_{JK}A^J_iA^K_j,\label{FYM}
\end{equation}
{}from which the Yang-Mills type Lagrangian is constructed as a quadratic invariant.

Specializing to the theory of gravity, we now identify the gauge symmetry group $G$ with the 10-parameter Poincar\'e group. As a semidirect product of the group of translations times the Lorentz group, it is conveniently parametrized by the set $\varepsilon^I = (\varepsilon^\alpha, \varepsilon^{\alpha\beta} = -\varepsilon^{\beta\alpha})$, hence we have the multi-index {\scriptsize{$I = \alpha,[\alpha\beta]$}}. The corresponding Poincar\'e gauge potentials
\begin{equation}\label{eG}
A^I_i = \left(\ e^\alpha_i,\quad \Gamma_i{}^{\alpha\beta} = -\Gamma_i{}^{\beta\alpha}\ \right)
\end{equation}
are then naturally interpreted as the coframe (tetrad) and the local Lorentz connection, respectively. They introduce the covariant derivative for the matter fields
\begin{equation}
D_\alpha\Phi^A = e^i_\alpha\left(\partial_i\Phi^A - {\frac 12}\Gamma_i{}^{\beta\gamma}(\rho_{\beta\gamma})^A_B\Phi^B\right).\label{DPP}
\end{equation}
Here $(\rho_{\beta\gamma})^A_B = -(\rho_{\gamma\beta})^A_B$ are the generators of the Lorentz transformations, and the factor 1/2 removes the double counting in the sum of the skew-symmetric objects.

The Poincar\'e gauge field strength tensors, using the Yang-Mills pattern (\ref{FYM}), read
\begin{eqnarray}
T_{ij}{}^\alpha &=& \partial_ie^\alpha_j - \partial_je^\alpha_i + \Gamma_{i\beta}{}^\alpha e^\beta_j - \Gamma_{j\beta}{}^\alpha e^\beta_i,\label{Tij}\\
R_{ij}{}^{\alpha\beta} &=& \partial_i\Gamma_j{}^{\alpha\beta}- \partial_j\Gamma_i{}^{\alpha\beta} + \Gamma_{i\gamma}{}^\beta\Gamma_j{}^{\alpha\gamma} - \Gamma_{j\gamma}{}^\beta\Gamma_i{}^{\alpha\gamma}.\nonumber\\
{}\label{Rij}
\end{eqnarray}
The anholonomic (Greek) indices are raised and lowered with the help of the Minkowski metric $g_{\alpha\beta}$. We identify the {\it translational} gauge field strength (\ref{Tij}) and the {\it rotational} gauge field strength (\ref{Rij}) with the torsion and curvature tensors, respectively. The first two terms on the right-hand side of (\ref{Tij}) form the anholonomity object $C_{ij}{}^\alpha = \partial_ie^\alpha_j - \partial_je^\alpha_i$. This is not a tensor under the local gauge group. The ``mixed'' form of (\ref{Tij}) is explained by the semidirect product (not direct product) structure of the Poincar\'e group.

In view of the skew symmetry of the connection, we can verify that the covariant derivative of the metric vanishes, $D_ig_{\alpha\beta} = 0$. One can solve the algebraic equation (\ref{Tij}) with respect to the connection (by cyclic permutation of indices) to find explicitly
\begin{equation}
\Gamma_{i\alpha\beta} = \tilde{\Gamma}_{i\alpha\beta} - K_{i\alpha\beta}.\label{GGK}
\end{equation}
Here the Riemannian connection is denoted by the tilde, and the post-Riemannian contortion tensor is determined by the torsion,
\begin{eqnarray}
\tilde{\Gamma}_{i\alpha\beta} &=& {\frac 12}\left(C_{\alpha\beta i} - C_{i\alpha\beta}
+ C_{i\beta\alpha}\right),\label{chr1}\\
K_{i\alpha\beta} &=& {\frac 12}\left(T_{\alpha\beta i} - T_{i\alpha\beta}
+ T_{i\beta\alpha}\right).\label{contortion}
\end{eqnarray}
Greek and Latin indices are converted into each other by means of the coframe: for example, $C_{\alpha\beta}{}^i = e^j_\alpha e^k_\beta e^i_\gamma C_{jk}{}^\gamma$. In particular, we thus find the components of the metric with respect to the local coordinate basis: $g_{ij} = e^\alpha_i e^\beta_j g_{\alpha\beta}$. Hence the spacetime interval
\begin{equation}
ds^2 = g_{ij}dx^i dx^j = g_{\alpha\beta}\vartheta^\alpha\vartheta^\beta\label{ds}
\end{equation}
is equivalently written either in terms of the holonomic coframe $dx^i$ or in terms of the anholonomic one $\vartheta^\alpha = e^\alpha_idx^i$.

%%%%%%%%%%%%%%%%%%%%%%%%%%%%%%%%%%%%%%%%%%%%%%%%%
\subsection{Dirac particle in Poincar\'e gravitational field}\label{DiracEq}
%%%%%%%%%%%%%%%%%%%%%%%%%%%%%%%%%%%%%%%%%%%%%%%%%

Let $\Psi$ be a Dirac spinor field. The corresponding generators of the Lorentz group are well known:
\begin{equation}
(\rho_{\alpha\beta}) = -\,{\frac i2}\sigma_{\alpha\beta},\qquad
\sigma_{\alpha\beta} = {\frac i2}\left(\gamma_\alpha \gamma_\beta
-\gamma_\beta\gamma_\alpha\right).\label{rhoL}
\end{equation}
The four Dirac matrices $\gamma^\alpha$, $\alpha = 0,1,2,3$, satisfy the standard anticommutation condition $\gamma^\alpha\gamma^\beta + \gamma^\beta\gamma^\alpha = 2g^{\alpha\beta}$. In addition to the local Poincar\'e symmetry, we assume the local $U(1)$ phase symmetry, which is responsible for the electromagnetic gauge field $A_i$. Accordingly, the total covariant derivative (\ref{DPP}) reads
\begin{equation}
D_\alpha\Psi = e_\alpha^i \left(\partial _i\Psi - {\frac {iq}{\hbar}}\,A_i\Psi
+ {\frac i4}\Gamma_i{}^{\beta\gamma}\sigma_{\beta\gamma}\Psi\right),\label{eqin2}
\end{equation}
where, making use of (\ref{DPhi}), we took into account that the generator of a 1-dimensional Abelian $U(1)$ group is $(\rho) = i$. The conventional $q/\hbar$ factor (where $q$ has the dimension of the electric charge) is needed to provide the correct dimension for the electromagnetic potential $A_i$ and for the (Maxwell) field strength $F_{ij} = \partial_iA_j - \partial_jA_i$.

The dynamics of a fermion particle with spin 1/2 and mass $m$ minimally coupled to the Poincar\'e gauge gravitational and electromagnetic field is described by the invariant action
\begin{equation}
I = \int\,d^4x\,{\cal L},\qquad {\cal L} = \sqrt{-g}\,L\label{action}
\end{equation}
where the Lagrangian reads
\begin{equation}\label{LD}
L = {\frac {i\hbar}{2}}\left(\overline{\Psi}\gamma^\alpha
D_\alpha\Psi - D_\alpha\overline{\Psi}\gamma^\alpha\Psi\right) -
mc\,\overline{\Psi}\Psi.
\end{equation}
The Dirac conjugate spinor is defined by $\overline{\Psi} = \Psi^\dagger\beta$ (with $\beta = c\gamma^{\hat{0}}$), and its covariant derivative reads
\begin{equation}
D_\alpha\overline{\Psi} = e_\alpha^i \left(\partial _i\overline{\Psi} + {\frac {iq}{\hbar}}
\,A_i\overline{\Psi} - {\frac i4}\Gamma_i{}^{\beta\gamma}\overline{\Psi}\sigma_{\beta\gamma}\right),\label{psibar}
\end{equation}

%%%%%%%%%%%%%%%%%%%%%%%%%%%%%%%%%%%%%%%%%%%%%%%%%
\subsection{Hermitian Hamiltonian for the Dirac fermion}\label{DiracH}
%%%%%%%%%%%%%%%%%%%%%%%%%%%%%%%%%%%%%%%%%%%%%%%%%

Let $x^i = (t,x^a)$ be the local coordinates on the spacetime manifold.

The study of the dynamics of the Dirac particle in an arbitrary Poincar\'e gauge field $(e^\alpha_i, \Gamma_i{}^{\alpha\beta})$ can be simplified for a convenient parametrization of the gravitational variables.

We describe the {\it translational gauge potential} (coframe $e^\alpha_i$) in the Schwinger \cite{Schwinger1,Schwinger2,dirac} gauge $e_a^{\,\widehat{0}} =0$ (also $e_{\widehat{a}}^{\,0} =0), ~a=1,2,3$, as follows:
\begin{equation}\label{coframe}
e_i^{\,\widehat{0}} = V\,\delta^{\,0}_i,\qquad e_i^{\widehat{a}} =
W^{\widehat a}{}_b\left(\delta^b_i - cK^b\,\delta^{\,0}_i\right),\qquad a=1,2,3.
\end{equation}
We assume that the functions $V$ and $K^a$, as well as the components of the $3\times 3$ matrix $W^{\widehat a}{}_b$ may depend arbitrarily on $t,x^a$.

One straightforwardly verifies that the coframe (\ref{coframe}) gives rise to a general form of the spacetime line element (\ref{ds})
\begin{equation}\label{LT}
ds^2 = V^2c^2dt^2 - \delta_{\widehat{a}\widehat{b}}W^{\widehat
a}{}_c W^{\widehat b}{}_d\,(dx^c - K^ccdt)\,(dx^d - K^dcdt).
\end{equation}
This is a slightly modified version of the well known parametrization of a metric proposed by Arnowitt, Deser, Misner \cite{ADM} and De Witt \cite{Dewitt} in the context of the canonical formulation of the quantum gravity theory; the off-diagonal metric components $g^{0a} = K^a/V^2c$ are related to the effects of rotation.

The components of {\it rotational gauge potential} (local Lorentz connection $ \Gamma_i{}^{\alpha\beta}$) are assumed to be completely arbitrary functions of $t,x^a$, too.

A direct check shows that the Schr\"odinger equation derived from the action (\ref{action}) has a non-Hermitian Hamiltonian. To avoid this difficulty, we define a new wave function by
\begin{equation}
\psi = \left({\frac 1c}{\sqrt{-g}e_{\widehat{0}}^0}\right)^{\frac
12}\,\Psi.\label{newpsi}
\end{equation}
Substituting the coframe (\ref{coframe}) into (\ref{eqin2}), (\ref{psibar}) and (\ref{action}), we rewrite the fermion action as
\begin{equation}\label{action2}
I = {\frac 12}\int\,dtd^3x\left[i\hbar\left(\psi^\dagger\partial_t\psi - \partial_t
\psi^\dagger\psi\right) - \psi^\dagger{\cal H}\psi + ({\cal H}\psi)^\dagger\psi\right].
\end{equation}
Here the {\it Hermitian} Hamiltonian reads 
\begin{eqnarray}
{\cal H} &=& \beta mc^2V + q\Phi + {\frac c 2}\left(\pi_b
\,{\cal F}^b{}_a \alpha^a + \alpha^a{\cal F}^b{}_a\pi_b\right)\nonumber\\
&& +\,{\frac c2}\left(\bm{K}\!\cdot\bm{\pi} + \bm{\pi}\!\cdot\!\bm{K}\right) +
{\frac {\hbar c}4}\left(\bm{\Xi}\!\cdot\!\bm{\Sigma} - \Upsilon\gamma_5\right),
\label{Hamilton1}
\end{eqnarray}
where the  kinetic 3-momentum operator $\pi_a = -i\hbar\partial_a  - qA_a = p_a - qA_a$ accounts of the interaction with the electromagnetic field $A_i = (-\Phi, A_a)$, and we denoted
\begin{eqnarray}
{\cal F}^b{}_a &=& VW^b{}_{\widehat a},\label{AB}\\ \label{ABU}
\Upsilon &=& V\epsilon^{\widehat{a}\widehat{b}\widehat{c}}\Gamma_{\widehat{a}\widehat{b}\widehat{c}},\\ 
\Xi^a &=& {\frac Vc}\,\epsilon^{\widehat{a}\widehat{b}\widehat{c}}\left(\Gamma_{\widehat{0}\widehat{b}\widehat{c}} + \Gamma_{\widehat{b}\widehat{c}\widehat{0}} + \Gamma_{\widehat{c}\widehat{0}\widehat{b}}\right).\label{ABX}
\end{eqnarray}
As usual,  $\alpha^a = \beta\gamma^a$ ($a,b,c,\dots = 1,2,3$) and the spin matrices
$\Sigma^1 = i\gamma^{\hat 2}\gamma^{\hat 3}, \Sigma^2 = i\gamma^{\hat 3}\gamma^{\hat 1},
\Sigma^3 = i\gamma^{\hat 1}\gamma^{\hat 2}$ and $\gamma_5=i\alpha^{\hat{1}}\alpha^{\hat{2}}
\alpha^{\hat{3}}$. Boldface notation is used for 3-vectors ${\bm K} = \{K^a\}, \,{\bm\pi}
= \{\pi_a\},\, {\bm\alpha} = \{\alpha^a\}, \,{\bm\Sigma} = \{\Sigma^a\}$. The three-dimensional totally antisymmetric Levi-Civita tensor $\epsilon^{\widehat{a}\widehat{b}\widehat{c}}$ has the only nontrivial component $\epsilon^{\widehat{1}\widehat{2}\widehat{3}} = 1$.

As a result, from the action (\ref{action2}) we derive the Schr\"odinger equation for the Dirac fermion particle in an arbitrary Poincar\'e gauge field  $(e^\alpha_i, \Gamma_i{}^{\alpha\beta})$:
\begin{equation}
i\hbar\frac{\partial \psi} {\partial t}= {\cal H}\psi.\label{Sch}
\end{equation}

%%%%%%%%%%%%%%%%%%%%%%%%%%%%%%%%%%%%%%%%%%%%%%%%%
\subsection{Spin-torsion coupling}\label{SpinTorsion}
%%%%%%%%%%%%%%%%%%%%%%%%%%%%%%%%%%%%%%%%%%%%%%%%%

In order to make the coupling of spin and torsion explicit, we now use the decomposition of the connection into the Riemannian and post-Riemannian parts (\ref{GGK})-(\ref{contortion}). Substituting (\ref{GGK}) into (\ref{ABU}) and (\ref{ABX}), we find
\begin{equation}\label{UX}
\Upsilon = \widetilde{\Upsilon} + Vc\check{T}^{\widehat{0}},\qquad
\Xi^{\widehat{a}} = \widetilde{\Xi}^{\widehat{a}} - V\check{T}^{\widehat{a}}.
\end{equation}
The tilde, as usual, denotes the Riemannian quantities
\begin{eqnarray}\label{ABU2}
\widetilde{\Upsilon} &=& V\epsilon^{\widehat{a}\widehat{b}\widehat{c}}\widetilde{\Gamma}_{\widehat{a}\widehat{b}\widehat{c}} = - V\epsilon^{\widehat{a}\widehat{b}\widehat{c}}{\cal C}_{\widehat{a}\widehat{b}\widehat{c}},\\ 
\widetilde{\Xi}_{\widehat{a}} &=& {\frac Vc}\,\epsilon_{\widehat{a}\widehat{b}\widehat{c}}\,\widetilde{\Gamma}_{\widehat{0}}{}^{\widehat{b}\widehat{c}} = \epsilon_{\widehat{a}\widehat{b}\widehat{c}}\,{\cal Q}^{\widehat{b}\widehat{c}},\label{ABX2}
\end{eqnarray}
which are constructed in terms of the following auxiliary objects:
\begin{eqnarray}
{\cal C}_{\widehat{a}\widehat{b}}{}^{\widehat{c}} &=& W^d{}_{\widehat{a}}
W^e{}_{\widehat{b}}\,\partial_{[d}W^{\widehat{c}}{}_{e]},\qquad {\cal
C}_{\widehat{a} \widehat{b}\widehat{c}} = g_{\widehat{c}\widehat{d}}
\,{\cal C}_{\widehat{a}\widehat{b}}{}^{\widehat{d}},\label{Cabc}\\
{\cal Q}_{\widehat{a}\widehat{b}} &=& g_{\widehat{a}\widehat{c}}W^d{}_{\widehat{b}}
\left({\frac 1c}\dot{W}^{\widehat c}{}_d + K^e\partial_e{W}^{\widehat c}{}_d
+ {W}^{\widehat c}{}_e\partial_dK^e\right).\nonumber\\
{}\label{Qab}
\end{eqnarray}
The dot $\dot{\,}$ denotes the derivative with respect to the coordinate time $t$. As we see, ${\cal C}_{\widehat{a}\widehat{b}}{}^{\widehat{c}} =- {\cal C}_{\widehat{b}\widehat{a}}{}^{\widehat{c}}$ is the reduced anholonomity object for the spatial triad ${W}^{\widehat a}{}_b$.

The non-Riemannian parts in (\ref{UX}) are constructed from the components of the axial torsion vector
\begin{equation}\label{Taxial}
\check{T}^\alpha = -\,{\frac 12}\,\eta^{\alpha\mu\nu\lambda}T_{\mu\nu\lambda},
\end{equation}
where $\eta^{\alpha\mu\nu\lambda}$ is the totally antisymmetric Levi-Civita tensor.

As a result, we can explicitly identify the spin-torsion coupling
\begin{equation}
-\,{\frac {\hbar cV}4}\left(\bm{\Sigma}
\!\cdot\!\check{\bm{T}} + c\gamma_5\check{T}{}^{\hat 0}\right),\label{ST}
\end{equation}
which comes from the last terms of the Dirac Hamiltonian (\ref{Hamilton1}). As usual,
$\check{\bm T} = \{\check{T}^a\}$.

%%%%%%%%%%%%%%%%%%%%%%%%%%%%%%%%%%%%%%%%%%%%%%%%%
\section{Foldy-Wouthuysen transformation for Dirac particle}\label{Hamiltonian}
%%%%%%%%%%%%%%%%%%%%%%%%%%%%%%%%%%%%%%%%%%%%%%%%%

At first, let us consider the purely gravitational case without the electromagnetic field. In order to reveal the physical contents of the Schr\"odinger equation (\ref{Sch}), we need to go to the Foldy-Wouthuysen (FW) representation. We can apply a general method for constructing the FW transformation developed in Ref. \cite{PRA} to the exact Dirac Hamiltonian (\ref{Hamilton1}).

Omitting the technical details (see \cite{OST,OSTRONG,ostgrav}) we then find for the
FW Hamiltonian:
\begin{equation}
{\cal H}_{FW}={\cal H}_{FW}^{(1)}+{\cal H}_{FW}^{(2)}+{\cal H}_{FW}^{(3)}.\label{eqFW}
\end{equation}
The three terms read, respectively,
\begin{widetext}
\begin{eqnarray}
{\cal H}_{FW}^{(1)} &=& \beta\epsilon' + \frac{\hbar c^2}{16}\left\{{\frac{1}
{\epsilon'}},\left(2\epsilon^{cae}\Pi_e \{p_b,{\cal F}^d{}_c\partial_d{\cal F}^b{}_a\}
+\Pi^a\{p_b,{\cal F}^b{}_a\widetilde{\Upsilon}\}\right)\right\} %\nonumber\\ && 
+ \frac{\hbar mc^4}{4}\epsilon^{cae}\Pi_e\left\{\frac{1}{{\cal T}},\left\{
p_d,{\cal F}^d{}_c{\cal F}^b{}_a\partial_bV\right\}\right\},\label{eq7}\\
% -----
{\cal H}_{FW}^{(2)} &=& \frac c2\left(K^a p_a + p_a K^a\right)
+ {\frac {\hbar c}4}\,\Sigma_a\widetilde{\Xi}^a\nonumber\\ 
&& +\,\frac{\hbar c^2}{16}\Biggl\{\frac{1}{{\cal T}},\biggl\{\Sigma_a
\{p_e,{\cal F}^e{}_b\},\Bigl\{p_f,\bigl[\epsilon^{abc}({\frac 1c} \dot{\cal
F}^f{}_c - {\cal F}^d{}_c\partial_dK^f + K^d\partial_d{\cal F}^f{}_c) %\nonumber\\ && 
- {\frac 12}{\cal F}^f{}_d\left(\delta^{db}\widetilde{\Xi}^a -
\delta^{da}\widetilde{\Xi}^b\right) \bigr]\Bigr\}\biggr\}\Biggr\},\label{eq7K}
\end{eqnarray}
\begin{eqnarray}
{\cal H}_{FW}^{(3)} &=& {\frac \hbar 2}\Sigma^a\Omega^{(T)}_a.\label{ST2}
\end{eqnarray}
Here the curly brackets $\{\ ,\ \}$ denote anticommutators and we introduced the operators
\begin{eqnarray}
\Omega^{(T)}_a &=& -\,{\frac {c}2}V\delta_{ab}\check{T}{}^{\hat b} + \beta{\frac
{c^3}{8}}\left\{\frac{1}{\epsilon'},\{p_b,{\cal F}^b{}_aV\check{T}^{\hat 0}\}\right\}
%\nonumber\\ && 
+\,\frac{c^2}{16}\Biggl\{\frac{1}{{\cal T}},\biggl\{\{p_e,{\cal F}^e{}_b\},
\Bigl\{p_f, {\cal F}^f{}_d V(\delta^{db}\check{T}{}^{\hat a} - \delta^{da}
\check{T}{}^{\hat b})\Bigr\}\biggr\}\Biggr\},\label{Lam}\\
\epsilon' &=& \sqrt{m^2c^4V^2+\frac{c^2}{4}\delta^{ac}\{p_b,{\cal F}^b{}_a\}
\{p_d,{\cal F}^d{}_c\}},\qquad {\cal T}=2{\epsilon'}^2 + \{\epsilon',mc^2V\}.\label{eqa}
\end{eqnarray}
The first two terms (\ref{eq7}) and (\ref{eq7K}) determine the dynamics of the Dirac fermion on the Riemannian spacetime manifold, whereas (\ref{ST2}) with (\ref{Lam}) gives the general description of the contribution of torsion field to the FW Hamiltonian. In the absence of torsion, we recover the previous results \cite{Ob1,Ob2,PRD,Warszawa,OST,OSTRONG,ostgrav}.
\end{widetext}

The equation of spin motion is obtained from the commutator of the
FW Hamiltonian with the polarization operator $\bm\Pi=\beta\bm\Sigma$:
\begin{equation}\label{spinmeq}
\frac{d\bm \Pi}{dt}=\frac{i}{\hbar}[{\cal H}_{FW},\bm{\Pi}] = \bm{\Omega}\times\bm{\Pi}.
\end{equation}
As a special case, let us consider the flat Minkowski metric with $V = 1, K^a = 0, W^{\widehat a}{}_b = \delta^a_b$. The spin precesses under the action of the torsion with the angular velocity $\bm{\Omega}= \bm{\Omega}^{(T)}$, where
\begin{eqnarray}
&& \bm{\Omega}^{(T)} = - \,{\frac{c}{2}}\bm{\check{T}} + \beta{\frac{c^3}{8}}\left\{
{\frac{1}{\epsilon'}},\left\{\bm{p},\check{T}^{\hat{0}}\right\}\right\}\nonumber\\ 
&& + \,{\frac c8}\left\{{\frac{c^2}{\epsilon'(\epsilon' + mc^2)}},\left(\left\{\bm{p}^2,\bm{\check{T}}\right\} - \left\{\bm{p},(\bm{p}\cdot\bm{\check{T}})\right\}\right)\right\}.\label{OmegaT0}
\end{eqnarray}
For slow nonrelativistic particles, this reduces to the earlier results of \cite{Rumpf,HS1,Aud,Bagrov1,Bagrov2,Nomura,ryder}.

%%%%%%%%%%%%%%%%%%%%%%%%%%%%%%%%%%%%%%%%%%%%%%%%%
\section{Nonminimal coupling: Covariant Dirac-Pauli equation}\label{DHamiltonian}
%%%%%%%%%%%%%%%%%%%%%%%%%%%%%%%%%%%%%%%%%%%%%%%%%

The conventional covariant Dirac equation disregards the anomalous magnetic moment and the electric dipole moment. Experimental search of the dipole moments of leptons and proton \cite{Semi,Commins,Mooser} (in particular in the study of physics beyond the Standard Model) calls forth the extensions of this equation, admitting a nonminimal coupling to the electromagnetic field. Taking into account the efforts to check the validity of the fundamental equivalence principle for particles with mass and spin (see Refs. \cite{Morgan,Haridass1,Haridass2,Ni77,Peres,Ritter1,Ritter2,Lamm,Mash00,Zhang,Kiefer,Teryaev:2003ch}, e.g.), it is necessary to investigate the possible nonminimal coupling to the gravitational field. For example, a possible violation of Einstein's equivalence principle can be manifest in the spin coupling to the Earth's rotation. In Ref. \cite{PRD2}, the bound on the anomalous gravitomagnetic moment (AGM) has been obtained from the re-analysis of the earlier experimental data. In order to develop a theoretical framework for the discussion of these issues, we consider here the covariant extension of the Dirac equation by going beyond the minimal coupling principle which is encoded in Eq. (\ref{eqin2}).

\subsection{Anomalous magnetic moment and electric dipole moment}\label{EDM}

In simple terms, the minimal coupling means that a gauge field enters the matter Lagrangian only via the covariant derivatives of the matter field. The nonminimal coupling is featured by the presence of explicit ``Pauli terms'' proportional to the gauge field strength. Let us discuss the electromagnetic interaction first. As one knows, the nonminimal term
\begin{equation}
{\frac{\mu'}{2c}}F_{ij}\overline{\Psi}\sigma^{ij}\Psi,\label{FSig}
\end{equation}
added to the Dirac Lagrangian (\ref{LD}), accounts for the possible anomalous magnetic moment (AMM) of a fermion particle coupled directly to the electromagnetic field strength tensor $F_{ij} = \partial_iA_j - \partial_jA_i$.

Noticing that $\sigma^{ij} = e_\alpha^i e_\beta^j\sigma^{\alpha\beta}$, we conclude that physically important are the anholonomic components of the field $F_{\alpha\beta}=e_\alpha^i e_\beta^jF_{ij}$. Introduce now the dual tensor by $G_{\alpha\beta} = {\frac 12}\eta_{\alpha\beta\mu\nu}F^{\mu\nu}$. The Lagrangian (\ref{LD}), modified by nonminimal coupling terms
\begin{equation}\label{DiracPauli}
{\frac {\mu'}{2c}}F_{\alpha\beta}\overline{\Psi}\sigma^{\alpha\beta}\Psi
+ {\frac {\delta'}2}G_{\alpha\beta}\overline{\Psi}\sigma^{\alpha\beta}\Psi
\end{equation}
describes the general case of a fermion with AMM and an electric dipole moment (EDM). The two coupling parameters have the dimension $[\mu'] = [q\hbar/2m]$ of the magnetic dipole (nuclear magneton), and $[\delta'] = [q\,l]$ of the electric dipole (charge times length), respectively.

Taking into account the nonminimal coupling (\ref{DiracPauli}), we find from (\ref{LD}) an extended Schr\"odinger equation with a modified Hamiltonian
\begin{eqnarray}
{\cal H} &=& \beta mc^2V + q\Phi + {\frac c 2}\left(\pi_b\,{\cal F}^b{}_a \alpha^a
+ \alpha^a{\cal F}^b{}_a\pi_b\right)\nonumber\\ 
&& + \,{\frac c2}\left(\bm{K}\!\cdot\bm{\pi} + \bm{\pi}\!\cdot\!\bm{K}\right) + {\frac {\hbar c}4}\left(\bm{\Xi}\!\cdot\!\bm{\Sigma} - \Upsilon\gamma_5\right)\nonumber\\ 
&& - \,\beta\left(\bm{\Sigma}\cdot\bm{\mathcal M} + i\bm{\alpha}\cdot\bm{\mathcal P}\right).\label{HamiltonDP}
\end{eqnarray}
Here we defined
\begin{eqnarray}
\bm{\mathcal M}^a &=& V\left(\mu'{\rm B}^a + \delta'{\rm E}^a\right),\label{Ma}\\
\bm{\mathcal P}_a &=& V\left(c\delta'{\rm B}_a - \mu'{\rm E}_a/c\right),\label{Pa}
\end{eqnarray}
in terms of the electric ${\rm E}_a = F_{\widehat{a}\widehat{0}}$ and magnetic ${\rm B}^a = {\frac 12}\,\epsilon^{\widehat{a}\widehat{b}\widehat{c}}F_{\widehat{b}\widehat{c}}$ fields (measured with respect to the anholonomic reference frame). 

\subsection{Gordon decomposition of Noether currents}\label{Gordon}

Before we turn to the analysis of the structure of the possible nonminimal coupling of a Dirac fermion to the Poincar\'e gauge field, it is instructive to recall the Gordon decomposition of the Noether currents \cite{Gordon28,seitz1,seitz2,lemke,Jad,Schuck}.

For the sake of maximal clarity, let us consider the dynamics of a free Dirac particle for which the Lagrangian (\ref{LD}) reduces to $L_D = {\frac {i\hbar}{2}}\left(\overline{\Psi}\gamma^\alpha e^i_\alpha\partial_i\Psi - \partial_i\overline{\Psi}\gamma^\alpha e^i_\alpha\Psi\right) - mc\,\overline{\Psi}\Psi$, with the trivial coframe $e^i_\alpha = \delta^i_\alpha$. This model, as it is well known, is invariant under the group $U(1)$ of the phase transformations of the wave function and under the Poincar\'e group of motion of the underlying flat Minkowski spacetime. These symmetries give rise, via the Noether theorem, to the three dynamical currents: the electromagnetic current, the canonical energy-momentum tensor, and the spin tensor, respectively,
\begin{eqnarray}
J^i &=& q\overline{\Psi}\gamma^i\Psi,\label{J1}\\
\Sigma_\alpha{}^i &=& {\frac {i\hbar}2}\left[\overline{\Psi}\gamma^i\partial_\alpha\Psi
- (\partial_\alpha\overline{\Psi})\gamma^i\Psi\right],\label{Dmom}\\
S_{\alpha\beta}{}^i &=& {\frac {\hbar}4}\overline{\Psi}(\gamma^i\sigma_{\alpha\beta} +
\sigma_{\alpha\beta}\gamma^i)\Psi.\label{Dspin}
\end{eqnarray}
These dynamical currents satisfy the conservation laws
\begin{equation}
\partial_iJ^i = 0,\qquad \partial_i\Sigma_\alpha{}^i = 0,\qquad
\partial_iS_{\alpha\beta}{}^i = \Sigma_{\alpha\beta} - \Sigma_{\beta\alpha}.\label{Dcons}
\end{equation}
The form of the last conservation law (of the total angular momentum) reflects the structure of the Poincar\'e group as a semidirect product of translations times the Lorentz group.

A remarkable feature of the Dirac dynamical currents is that one can decompose them into two pieces, namely, into the convective and polarizational parts as follows:
\begin{eqnarray}
J^i &=& {\stackrel c J}{}^i + \partial_jM^{ij},\label{GD1}\\
\Sigma_\alpha{}^i &=& {\stackrel c \Sigma}{}_\alpha{}^i + \partial_j
\check{M}_\alpha{}^{ij},\label{GD2}\\
S_{\alpha\beta}{}^i &=& {\stackrel c S}{}_{\alpha\beta}{}^i + \partial_jM_{\alpha\beta}{}^{ij}
+ \check{M}_{\alpha\beta}{}^i - \check{M}_{\beta\alpha}{}^i.\label{GD3}
\end{eqnarray}
For the electromagnetic current (\ref{GD1}) this was noticed by Gordon \cite{Gordon28} shortly after Dirac established his relativistic wave equation for a spin 1/2 particle, and later \cite{seitz1,seitz2,lemke,Jad,Schuck} this decomposition was demonstrated for the particles of any spin and generalized for the gravitational currents (\ref{GD2}) and (\ref{GD3}).

The convective parts ${\stackrel c J}{}^i, {\stackrel c \Sigma}{}_\alpha{}^i, {\stackrel c S}{}_{\alpha\beta}{}^i$ turn out to be the Noether currents (corresponding to $U(1)$ and Poincar\'e symmetries) for the convective Lagrangian $L_C = {\frac {\hbar^2}{2mc}}\partial^j\overline{\Psi}\partial_j\Psi - {\frac {mc}2}\overline{\Psi}\Psi$. It is worthwhile to notice that the field equation for $L_C$ coincides with the squared Dirac equation $\square\Psi - {\frac {m^2c^2}{\hbar^2}}\Psi = 0$.

The polarizational currents (\ref{GD1})-(\ref{GD3}) are expressed in terms of the {\it dipole moments}
\begin{eqnarray}
M^{ij} &=& {\frac {q\hbar}{2mc}}\overline{\Psi}\sigma^{ij}\Psi,\label{M1}\\
M_{\alpha}{}^{ij} &=& {\frac {i\hbar^2}{4mc}}\left(\overline{\Psi}\sigma^{ij}\partial_\alpha\Psi -\partial_\alpha\overline{\Psi}\sigma^{ij}\Psi\right),\label{M2}\\
M_{\alpha\beta}{}^{ij} &=& {\frac {\hbar^2}{8mc}}\left(\overline{\Psi}\sigma^{ij}\sigma_{\alpha\beta}\Psi + \overline{\Psi}\sigma_{\alpha\beta}\sigma^{ij}\Psi\right),\label{M3}
\end{eqnarray}
and the modified moment $\check{M}_{\alpha}{}^{ij} = M_{\alpha}{}^{ij} + e^i_\alpha M_{k}{}^{jk} - e^j_\alpha M_{k}{}^{ik}$. The complex structure of the Gordon decompositions (\ref{GD2}) and (\ref{GD3}) is again related to the semidirect product nature of the Poincar\'e group.

The physical interpretation of the moments (\ref{M1})-(\ref{M3}) is crystal clear: these are Amp\'ere dipoles generated by the matter currents \cite{Chou} that carry electric charge, gravitational translational charge (mass) and gravitational rotational charge (spin), respectively. As we can see, the corresponding generators of $U(1)$ and Poincar\'e groups explicitly determine the structure of the respective dipole moment.

One can generalize these observations to the curved spacetime \cite{seitz1,seitz2,lemke,Jad,Schuck}. Qualitatively, this amounts to the replacement of the partial derivatives by the covariant ones in the equations above. However, the important point is that the form of the dipole moments (\ref{M1})-(\ref{M3}) remain the same for any spacetime geometry.

\subsection{Poincar\'e gravitational moments}\label{Gravmoments}

We are now in position to discuss the possible form of the nonminimal coupling of a Dirac fermion to the Poincar\'e gauge gravitational field. The key is provided by the Pauli term (\ref{FSig}) which has the transparent structure of a product of the electromagnetic field strength times the electromagnetic moment (\ref{M1}): $\sim F_{ij}M^{ij}$.

Therefore, taking into account the existence of a dipole moment (\ref{M1})-(\ref{M3}) for every symmetry generator, we come to the natural conclusion that possible nonminimal coupling to the Poincar\'e gauge follows the same electrodynamical pattern. Namely, the corresponding gravitational Pauli type terms have the same product structure of the translational gauge field strength (\ref{Tij}) times the translational moment (\ref{M2}) plus the rotational gauge field strength (\ref{Rij}) times the rotational moment (\ref{M3}): $\sim T_{ij}{}^\alpha M_\alpha{}^{ij} + R_{ij}{}^{\alpha\beta} M_{\alpha\beta}{}^{ij}$.

Explicitly, in addition to (\ref{FSig}) the possible covariant gravitational nonminimal coupling terms read as follows
\begin{eqnarray}
&& {\frac {\rho'}{2}}\,T_{ij}{}^\alpha \left(\overline{\Psi}\sigma^{ij}D_\alpha\Psi - D_\alpha\overline{\Psi}\sigma^{ij}\Psi\right)\nonumber\\ 
&& + {\frac {\tau'}{2}}\,R_{ij}{}^{\alpha\beta}\overline{\Psi}\left(\sigma^{ij}\sigma_{\alpha\beta} + \sigma_{\alpha\beta}\sigma^{ij}\right)\Psi.\label{TMRM}
\end{eqnarray}
The two new coupling parameters have the same physical dimension $[\rho'] = [\tau'] = [\hbar l]$ (spin times length).

We can simplify the second term in (\ref{TMRM}) by making use of the Dirac algebra to
\begin{equation}
-\,\tau' R\,\overline{\Psi}\Psi + i\tau'P\,\overline{\Psi}\gamma_5\Psi,
\end{equation}
where $R = R_{ij}{}^{\alpha\beta}e^j_\alpha e^i_\beta$ is the curvature scalar, and $P = R_{\alpha\beta\gamma\delta}\eta^{\alpha\beta\gamma\delta}$ is the pseudoscalar of the Riemann-Cartan curvature. When the torsion is zero, the nonminimal coupling (\ref{TMRM}) reduces to $-\tau' \tilde{R}\,\overline{\Psi}\Psi$. This is a typical curvature dependent term which arises naturally in the squared Dirac equation.

It is worthwhile to mention that the field-theoretic models with the nonminimal coupling of the type (\ref{TMRM}) were discussed not only in the framework of the Poincar\'e gauge theory \cite{Shapiro,Haridass2} but also in the context of the search of the possible signatures of the Lorentz-violating effects \cite{Tasson,Kost,Bonder}.

\subsection{On gravitomagnetic and gravitoelectric moments}\label{GMEMs}

There is a formal analogy between gravitational and electromagnetic phenomena known as the gravitoelectromagnetism \cite{gem1,gem2,gem3} that can be established for the weak gravitational fields. 

In the Riemannian framework of Einstein's GR (no torsion), it was observed in Ref. \cite{OSTRONG} that the squared Dirac equation features -- in the weak field approximation -- a common effect that is produced on the spin by the electromagnetic and the gravitational (or inertial) fields via the term $\sigma^{\alpha\beta}\left(qF_{\alpha\beta}/c + m\Phi_{\alpha\beta}\right)$, where $\Phi_{\alpha\beta}= \left\{\pi^{i}, \Gamma_{i\,\alpha\beta}\right\}/(2m)$. 

In the semiclassical approximation, momentum is proportional to the velocity, $\pi^{i}=mU^{i}$, and $\Phi_{\alpha\beta}$ coincides with the spin transport matrix in a gravitational field (see \cite{OST} and \cite{PK}). Making this observation, one could expect that a Dirac particle may have a nontrivial gravitomagnetic moment along with the magnetic moment. The theoretical analysis \cite{PRD2} established a strong bound on the anomalous gravitomagnetic moment from experiment \cite{Venema}.

However, such weak-field considerations are not covariant. The electromagnetic field strength $F_{ij}$ is tensor and hence $\sigma^{\alpha\beta}F_{\alpha\beta}=\sigma^{ij}F_{ij}$ is invariant under arbitrary coordinate and Lorentz frame transformations. In contrast, the local Lorentz connection $\Gamma_{i\,\alpha\beta}$ is not a tensor, hence $\sigma^{\alpha\beta}\Phi_{\alpha\beta}$ is not an invariant object. An attempt to extend $\sigma^{\alpha\beta}\Phi_{\alpha\beta}$ via the identity \cite{OSTRONG}
\begin{eqnarray}
&& g^{ij}(\hbar^2D_iD_j + \pi_{i}\pi_{j}) = {\frac {\hbar m}{2}}\left[\sigma^{\alpha\beta}
\Phi_{\alpha\beta}\right.\nonumber\\
&& \left. -\frac{\hbar}{8m}\left(2\Gamma^i{}_{\alpha\beta}\Gamma_i{}^{\alpha\beta} + i\gamma_5\eta^{\alpha\beta\mu\nu}\Gamma^i{}_{\alpha\beta}\Gamma_{i\,\mu\nu}\right)\right]\label{relation}
\end{eqnarray}
also fails since both sides are non-covariant in view of the non-covariance of $\pi_{i}$.

We thus conclude that the anomalous gravitomagnetic moment is not allowed in the covariant Dirac-Pauli theory with a nonminimal coupling of a fermion to the Poincar\'e gauge gravitational field. This demonstrates a limited nature of analogies between gravitational and electromagnetic interactions observed in the weak-field approximation. The same conclusion is valid for the anomalous gravitoelectric moment. It is worthwhile to recall that the analysis of the gravitational form-factors of Dirac fermions by Kobzarev and Okun \cite{KO} (see also \cite{Leitner,Hiida}) have shown that the anomalous gravitomagnetic and gravitoelectric moments should be strictly zero.

For the Riemann-Cartan geometry, the terms (\ref{TMRM}) open a different possibility for a nonminimal coupling of the Poincar\'e gauge gravitational field with the gravitational moments of a fermion particle.

%%%%%%%%%%%%%%%%%%%%%%%%%%%%%%%%%%%%%%%%%%%%%%%%%
\section{Spin-1/2 particle in magnetic field and rotating frame}\label{MFRF}
%%%%%%%%%%%%%%%%%%%%%%%%%%%%%%%%%%%%%%%%%%%%%%%%%

In the next section we will estimate the possible effects of the spacetime torsion on the basis of the experimental data for the cold neutrons and atoms affected by the gravitational field of the rotating Earth. Here we provide the necessary theoretical framework for this analysis. The reference frame rotating with the angular velocity $\bm\omega$ is given by
\cite{HN}
\begin{equation}
V = 1,\qquad W^{\widehat a}{}_b = \delta^a_b,\qquad K^a =
-\,{\frac {(\bm{\omega}\times\bm{r})^a}{c}}.\label{VWni}
\end{equation}

Substituting this into the Hamiltonian (\ref{Hamilton1}), we find the Schr\"odinger description of a Dirac particle in the uniform magnetic field $\bm B$ and rotating frame
\begin{equation}\label{Hamlroton}
{\cal H} = \beta mc^2 + c\bm{\alpha}\cdot\bm{\pi} - \bm{\omega}\cdot\bm{\lambda}
- {\frac{\hbar}{2}}\bm{\omega}\cdot\bm{\Sigma} - {\frac{\hbar c}{4}}\left(
\check{T}^{\hat{0}} c\gamma_5 + \bm{\check{T}}\cdot\bm{\Sigma}\right).
\end{equation}
Here $\bm\lambda=\bm r\times\bm\pi=-\bm\pi\times\bm r$ denotes the orbital angular moment operator, and the torsion effects are encoded in the last term.

Applying the FW transformation to the Hamiltonian (\ref{Hamlroton}), we find
\begin{equation}
{\cal H}_{FW}={\cal H}_{0}+{\cal H}_{1},\label{Hamlt}
\end{equation}
\begin{widetext}
where
\begin{eqnarray}
{\cal H}_{0} &=& \beta\epsilon - \bm{\omega}\cdot\bm{\lambda} - {\frac \hbar2}
\bm{\omega}\cdot\bm{\Sigma} + {\frac{\hbar c^3}{16}}\left\{{\frac{1}{\epsilon}},
\left\{\bm{\pi}\cdot\bm{\Pi},\check{T}^{\hat{0}}\right\}\right\} -
{\frac{\hbar c}{4}}\bm{\check{T}}\cdot\bm{\Sigma}\nonumber\\
&& +\,{\frac{\hbar c}{16}}\left\{{\frac{c^2}{\epsilon(\epsilon+mc^2)}},\left[
\left\{\bm{\pi}^2,\bm{\check{T}}\cdot\bm{\Sigma}\right\} - {\frac{1}{2}}\left\{
\bm{\Sigma}\cdot\bm{\pi},(\bm{\pi}\cdot\bm{\check{T}} + \bm{\check{T}}\cdot
\bm{\pi})\right\}\right]\right\},\label{HFW0}\\
{\cal H}_{1} &=& -\,{\frac{e\hbar c}{8}}\left\{{\frac{1}{\epsilon(\epsilon+mc^2)}},
\bm{\Sigma}\cdot(\bm{G}\times\bm{\pi} - \bm{\pi}\times\bm{G})\right\}.\label{HFW1}
\end{eqnarray}
Here we take into account that $q=e$ and denote
\begin{equation}
\epsilon=\sqrt{m^2c^4+c^2\bm{\pi}^2 - e\hbar c^2\bm{\Sigma}\cdot\bm{B}},
\qquad\bm{G} = \bm{B}\times(\bm{\omega}\times\bm{r}).\label{eptau}
\end{equation}
%\end{widetext}
In what follows we identify $\bm{\omega}$ with the angular velocity of the Earth. Evidently, $\bm{\omega}\times\bm{r}$ is the particle velocity in the inertial system related to the centre of the Earth. Note that $\bm{G} = \left\{{\rm E}_a\right\}$ is the electric field in a rotating frame. Eq.~(\ref{HFW1}) describes the main correction to the Hamiltonian. Next-to-leading order corrections to ${\cal H}_{0}$ are of order of $\hbar^2$ and, moreover, they do not depend on spin. ${\cal H}_{1}$ is much less than ${\cal H}_{0}$ and can be neglected since the kinetic momentum $\bm\pi$ is usually zero on average. The corrections to ${\cal H}_{0}$ are of the same order for Dirac particles both in uniform and nonuniform magnetic fields.

For the actual experimental conditions we have $|e\hbar B|\ll m^2c^2$ in (\ref{HFW1}), that is the magnetic field is much smaller than the critical field $|B|\ll B_c = m^2c^2/e\hbar$. This allow us to take into account only terms linear in the magnetic field. In this approximation,
\begin{equation}
\epsilon = \epsilon' - \left\{{\frac{e\hbar c^2}{2\epsilon'}},\bm{\Sigma}
\cdot\bm{B}\right\} = \epsilon' - \left\{{\frac{\mu_0}{\gamma}},\bm{\Sigma}\cdot
\bm{B}\right\}, \qquad \epsilon' = \sqrt{m^2c^4 + c^2\bm{\pi}^2},\label{eps1}
\end{equation}
where $\mu_0 = {\frac {e\hbar}{2m}}$ is the Dirac magnetic moment and $\gamma = {\frac {\epsilon'}{mc^2}}$ is the Lorentz factor.

Let us now consider the spin dynamics described by the precession equation (\ref{spinmeq}). Using the FW Hamiltonian (\ref{Hamlt}), we then find the corresponding operator of the angular velocity of the spin rotation:
\begin{eqnarray}
\bm{\Omega} = -\,\beta\left\{{\frac{2\mu_0}{\hbar\gamma}},\bm{B}\right\} - \bm{\omega}
- {\frac{c}{2}}\bm{\check{T}} + \beta{\frac{c^3}{8}}\left\{{\frac{1}{\epsilon'}},
\left\{\bm{\pi},\check{T}^{\hat{0}}\right\}\right\} %\nonumber\\ && 
+ {\frac{c}{8}}\left\{{\frac{c^2}{\epsilon'(\epsilon'+mc^2)}},\left[\left\{
\bm{\pi}^2,\bm{\check{T}}\right\} % \right.\right.\nonumber\\ && \left.\left. 
- {\frac{1}{2}}\left\{\bm{\pi},(\bm{\pi}\cdot
\bm{\check{T}} + \bm{\check{T}}\cdot\bm{\pi})\right\}\right]\right\}.\label{Omega}
\end{eqnarray}
Here the contribution of ${\cal H}_{1}$ is neglected. The resulting expression is sufficiently precise for realistic magnetic field in the actual experiment.

For a more general case of a fermion spin-1/2 particle with a nontrivial magnetic moment ($\mu'\neq 0$ and $\delta'= 0$) in a magnetic field and in rotating frame the dynamics is described by the Hermitian Hamiltonian (\ref{HamiltonDP}). It reads explicitly
\begin{eqnarray}
{\cal H} = \beta mc^2 + c\bm{\alpha}\cdot\bm{\pi} - \bm{\omega}\cdot\bm{\lambda}
- {\frac{\hbar}{2}}\bm{\omega}\cdot\bm{\Sigma} - \mu'\bm{\Pi}\cdot\bm{B} %\nonumber\\ &&
- {\frac{\hbar c}{4}}\left(\check{T}^{\hat{0}}c\gamma_5 
+ \bm{\check{T}}\cdot\bm{\Sigma}\right).\label{Hamilton2}
\end{eqnarray}
As compared to Eq. (\ref{Hamlroton}), this equation includes the contribution of the AMM.

When the magnetic field is uniform, the FW transformation of (\ref{Hamilton2}) results in
\begin{equation}
{\cal H}_{FW}={\cal H}_{0}+{\cal H}_{1}+{\cal H}_{2},\label{Hamiltn}
\end{equation}
%\begin{widetext}
where ${\cal H}_{0}$ and ${\cal H}_{1}$ are defined by (\ref{HFW0})
and (\ref{HFW1}), whereas the last term is equal to
\begin{equation}
{\cal H}_{2} = -\,\mu'\bm{\Pi}\cdot\bm{B} + {\frac{\mu'}{4}}\left\{{\frac{c^2}
{\epsilon'(\epsilon' + mc^2)}},\left[(\bm{B}\cdot\bm{\pi})(\bm{\Pi}\cdot\bm{\pi})
+ (\bm{\Pi}\cdot\bm{\pi})(\bm{\pi}\cdot\bm{B})\right]\right\}.\label{HFW2}
\end{equation}
Taking this term into account, the angular velocity of spin
rotation (\ref{Omega}) is modified:
\begin{eqnarray}
\bm{\Omega} &=& -\,\bm{\omega} -\beta\left\{{\frac{\mu_0 mc^2}{\hbar\epsilon'}},
\bm{B}\right\} - 2\beta{\frac{\mu'}{\hbar}}\bm{B} + \frac{\mu'}{2\hbar}\left\{
\frac{c^2}{\epsilon'(\epsilon'+mc^2)},\bigl[(\bm{B}\!\cdot\!\bm{\pi})\bm{\pi}
+ \bm{\pi}(\bm{\pi}\!\cdot\!\bm{B})\bigr]\right\}\nonumber\\
&& -\,{\frac{c}{2}}\bm{\check{T}} + \beta{\frac{c^3}{8}}\left\{{\frac{1}{\epsilon'}},
\left\{\bm{\pi},\check{T}^{\hat{0}}\right\}\right\} + {\frac{c}{8}}\left\{
{\frac{c^2}{\epsilon'(\epsilon'+mc^2)}},\left[\left\{\bm{\pi}^2,\bm{\check{T}}
\right\} - (\bm{\check{T}}\cdot\bm{\pi})\bm{\pi} - \bm{\pi}(\bm{\pi}\cdot
\bm{\check{T}})\right]\right\}.\label{Omegaam}
\end{eqnarray}
Evaluating the anticommutators in Eqs. (\ref{HFW2}) and (\ref{Omegaam}), we can find the effects of a possible non-uniformity of the magnetic field.
\end{widetext}

%%%%%%%%%%%%%%%%%%%%%%%%%%%%%%%%%%%%%%%%%%%%%%%%%
\section{Experimental bounds on spin-torsion coupling}\label{NB}
%%%%%%%%%%%%%%%%%%%%%%%%%%%%%%%%%%%%%%%%%%%%%%%%%

The theoretical analysis of the dynamics of spin underlie the discussion of possible verifications of the Poincar\'e gauge gravity \cite{Hehl71,Adamowicz,AudLam,HS2,HS3,AudLNP,Ivanov}, see also \cite{Shapiro,Rumpf,HS1,Aud}. As compared to the extensive theoretical research, only few experimental studies were directly devoted to the search of the spin-torsion coupling \cite{WeiTou,Ni2010,Lehnert}. However we can use the theoretical framework established in our paper to find observational bounds on spin-torsion coupling from the experimental data available in the literature.

In a large class of experiments, the dynamics of freely precessing nuclear spins in a uniform magnetic field was investigated by making use of comagnetometers with two different kinds of atoms in $S$-states. Ratios of their nuclear $g$-factors were either defined with a needed precision or measured during an experiment. More specifically, the relevant measurements were reported \cite{Venema} for the experiment with $^{199}$Hg and $^{201}$Hg atoms devoted to the search of a hypothetical scalar-pseudoscalar interactions. The atoms were at rest. The experimental data \cite{Venema} was earlier used in \cite{PRD2} to derive estimates for the anomalous gravitomagnetic moment. Here we exclude the latter from our consideration since the the anomalous gravitomagnetic moment cannot be introduced in a fully consistent covariant way. To determine bounds on the spacetime torsion, we also disregard the scalar-pseudoscalar interactions and present the spin-dependent part of the FW Hamiltonian as follows:
\begin{equation}\label{nonrelt}
{\cal H}_{FW} = -\,(\mu_0+\mu')\bm{B}\cdot\bm{\Pi} - {\frac\hbar2}\bm{\omega}
\cdot\bm{\Sigma} - {\frac{\hbar c}{4}}\bm{\check{T}}\cdot\bm{\Sigma}.
\end{equation}

It has been demonstrated in \cite{JINRLett12} that the classical limit of relativistic FW Hamiltonians can be obtained by a simple replacement of quantum mechanical operators with corresponding classical quantities. Therefore, the classical limit of Hamiltonian (\ref{nonrelt}) reads
\begin{equation}
H = -\,g_N\frac{\mu_N}{\hbar}\bm{B}\cdot\bm{s} - \bm{\omega}\cdot\bm{s}
- {\frac{c}{2}}\bm{\check{T}}\cdot\bm{s},\label{nonclas}
\end{equation} where $g_N$ is the nuclear $g$-factor and $\mu_N$ is the nuclear magneton.

Let us denote two kinds of atoms by the subscripts 1 and 2. The measured ratio of Zeeman frequencies for transitions between neighbouring atomic levels, $R=\nu_{2}/\nu_{1}$, depends on the direction of the magnetic field $\bm{B}$ and on the spin-torsion coupling. Two opposite directions of the magnetic field were used in experiment \cite{Venema}. The calculation of the difference of these ratios for the two opposite directions (labeled by $\pm$ subscripts below) of magnetic field is similar to the derivations done in Ref. \cite{PRD2} and the result reads
\begin{eqnarray}
R_+-R_- = \pm\frac {1-{\cal G}}{2\pi\nu_1}\left[2\omega\cos\theta 
+ c |\bm{\check{T}}|\cos\Theta\right].\label{sol}
\end{eqnarray}
Here ${\cal G} =\frac{g_2}{g_1}$ is the ratio of $g-$factors; $\theta$ is the angle between the direction of magnetic field $\bm B$ and the Earth's rotation axis, whereas $\Theta$ is the angle between $\bm{B}$ and the torsion $\bm{\check{T}}$; $\omega$ is the magnitude of the Earth's angular velocity, and $\nu_1$ is the Zeeman frequency for atoms of the first kind. The experimental conditions of \cite{Venema} for $^{199}$Hg and $^{201}$Hg atoms correspond to the angle $\theta\approx 0$, and the ratio of $g$-factors is ${\cal G}=-0.369139$. Using the experimental data from \cite{Venema}, we then obtain the restriction on the absolute value of the spacetime torsion:
\begin{eqnarray}
{\frac{\hbar c}{4}}|\,\bm{\check{T}}|\cdot|\cos\Theta| < 2.2\times10^{-21}\,{\rm eV},
\nonumber\\ 
|\bm{\check{T}}|\cdot|\cos\Theta| < 4.3\times10^{-14}\,{\rm m}^{-1}. \label{restrctn}
\end{eqnarray}

In the same manner, we can re-analyze the similar experiments \cite{Gemmel,Gemmel2,burghoff,Fab} where the difference of the weighted Zeeman frequencies was measured for He and Xe atoms:
\begin{eqnarray}
|\Delta\nu| = |\nu_2 - {\cal G}\nu_1| = \left|{\frac {1-{\cal G}}{2\pi}}\left(\omega\cos\theta + {\frac{c}{2}}|\bm{\check{T}}|\cos\Theta\right)\right|.
\nonumber\\ 
%{\cal G} =\frac{g_2}{g_1}
\label{Gem}
\end{eqnarray}
Making use of the experimental data presented in Sec. 4.3 of Ref. \cite{Gemmel}, we can extract the new restriction on the minimal coupling of torsion (with the $g$-factor ratio ${\cal G} =g_{He}/g_{Xe} = 2.75408159(20)$, and $(1-{\cal G})\omega\cos\theta = -\,6.87263 \times10^{-5}$ rad/s):
\begin{eqnarray}
{\frac{\hbar c}{2}}|\,\bm{\check{T}}|\cdot|(1-{\cal G})\cos\Theta| < 4.1\times10^{-22}\,{\rm eV},\nonumber\\ 
|\bm{\check{T}}|\cdot|\cos\Theta|< 2.4\times10^{-15}\,{\rm m}^{-1}.\label{Gemresults}
\end{eqnarray}
Equations (\ref{restrctn}) and  (\ref{Gemresults}) present the strong new bounds on the spacetime torsion. 

\section{Discussion}\label{Conclusion}

In this paper, we have studied the dynamics of the Dirac fermion particle in the framework of the Poincar\'e gauge gravity theory. This problem is of considerable interest because one cannot probe the possible deviations of the spacetime structure from the Riemannian geometry with the help of the spinless matter (massive test particles or extended test bodies) even if the latter is characterized by a macroscopic angular momentum. Only matter with intrinsic spin is affected by the spacetime torsion \cite{YS,PO2007}, and in this sense a Dirac fermion appears to be a natural measuring device for the torsion experiments.

The quantum dynamics of the spin-$1/2$ particle minimally coupled to an arbitrary Poincar\'e  gauge field $(e^\alpha_i, \Gamma_i{}^{\alpha\beta})$ was analysed in detail in Secs.~\ref{DiracEq}-\ref{Hamiltonian} and the Foldy-Wouthuysen Hamiltonian was derived with no assumptions about the weakness of the fields. This central result underlies the subsequent study of the behaviour of the spin under the influence of the external fields (electromagnetic, inertial, Riemannian gravitational and non-Riemannian torsion).

Possible covariant extensions of the Dirac theory to the nonminimal Pauli-type coupling were discussed in Sec.~\ref{DHamiltonian}, where the important role of the gravitational moments (translational and Lorentz) was clarified. They are introduced on the basis of the fundamental Gordon decomposition technique of the Noether currents \cite{Gordon28,seitz1,seitz2,lemke,Jad,Schuck}. These gravitational moments (together with their Hodge duals) provide a regular way to construct a consistent covariant theory of a Dirac fermion particle with an intrinsic dipole structure induced by the physical Noether charges.

It is worthwhile to mention that the analysis in Sec. \ref{GMEMs} proved that the anomalous gravitomagnetic and gravitoelectric moments cannot be introduced in a covariant way for Dirac fermions. The earlier results of Kobzarev and Okun \cite{KO} relate the validity of the equivalence principle to the absence of both the anomalous gravitomagnetic and the gravitoelectric dipole moment, defined as the formal gravitational analogs of the anomalous magnetic moment and the electric dipole moment, respectively. This important point is apparently under-appreciated in the literature (for example, it is not mentioned in the nice recent review \cite{Ni2010}). Relations obtained by Kobzarev and Okun predict equal frequencies of the precession of all classical and quantum spins in any curved spacetimes \cite{Teryaev:2003ch}. In the weak-field approximation, the analysis \cite{PRD2} of the earlier experimental data has put a bound of about 4$\%$ on the anomalous gravitomagnetic moment. As mentioned in Ref. \cite{Ni2010}, the experimental data by Kornack \emph{et al} \cite{Kornack} give the restriction of 3$\%$, whereas a stronger restriction of 0.9$\%$ has been obtained in Ref. \cite{OST2014} on the basis of the experimental data of Ref. \cite{Gemmel}.

In Sec.~\ref{NB}, we have established new strong bounds on the possible background spacetime torsion for the minimally coupled Dirac fermion. The results obtained are consistent with the earlier estimates of the torsion derived from the Hughes-Drever type experiments \cite{Laemm}, and with the experimental limits found in the framework of the search of the Lorentz symmetry violations \cite{Kost2}.

%%%%%%%%%%%%%%%%%%%%%%%%%%%%%%%%%%%%%%%%%%%%%%%%%
\section*{Acknowledgements}
%%%%%%%%%%%%%%%%%%%%%%%%%%%%%%%%%%%%%%%%%%%%%%%%%

We thank Friedrich Hehl for reading the preliminary draft and for his many valuable comments. A.S. and O.T. are also grateful to S. Karpuk and F. Allmendinger for stimulating discussions. This work was supported in part by the Belarusian Republican Foundation for Fundamental Research (Grant No. $\Phi$14D-007), by the program of collaboration between Bogoliubov Laboratory of Theoretical Physics (JINR) and Belarus, by the Deutsche Forschungsgemeinschaft (Grant No. 436 RUS 113/881/0), by the Heisenberg-Landau program of the German Ministry for Science and Technology (BMBF), and by the Russian Foundation for Basic Research (Grants No. 12-02-00613 and No. 14-01-00647).

\appendix

\section{Conventions \& Symbols}\label{conventions_app}

\begin{table}
\caption{\label{tab_symbols}Directory of symbols.}
\begin{ruledtabular}
\begin{tabular}{ll}
Symbol & Explanation\\
\hline
%&\\
\hline
\multicolumn{2}{l}{{Spacetime geometry}}\\
\hline
$g_{\alpha\beta}$, $g_{ij}$ & Metric\\
$\sqrt{-g}$ & Determinant of the metric \\
$\delta^a_b$ & Kronecker symbol \\
$\eta_{\alpha\beta\mu\nu}$ & Levi-Civita tensor \\
$x^i = (t, x^{a})$ & Coordinates (time, space) \\
$dx^i$, $\vartheta^\alpha$ & Coframe one-form \\
$e_i^\alpha$ & Tetrad \\ 
$\Gamma_{i\beta}{}^\alpha$ & Connection \\
$K_{i\beta}{}^\alpha$ & Contortion \\
$C_{ij}{}^\alpha$ & Anholonomity object \\
$T_{ij}{}^\alpha$ & Torsion \\
$R_{ij\beta}{}^\alpha$& Curvature \\
$\check{T}{}^\alpha$, $\check{T}{}^{\hat{0}}$, $\check{\bm T}$ & Axial torsion\\
$V$, $W^{\hat{a}}{}_b$, $K^a$, ${\cal F}^b{}_a$ & Metric constituents \\
$\Upsilon$, $\Xi^a$, ${\cal Q}_{\hat{a}\hat{b}}$, ${\cal C}_{\hat{a}\hat{b}}{}^{\hat{c}}$ 
& Connection constituents\\
$\partial_i$, $D_i$, $D_\alpha$ & (Partial, covariant) derivative \\ 
\hline
\multicolumn{2}{l}{{Matter and gauge fields}}\\
\hline
$\Phi^A$ & General matter field \\
$A^I_i$ & Gauge field (potential) \\
$F_{ij}{}^I$ & Gauge field strength \\
$A_i$, $F_{ij}$, ${\rm E}_a$, ${\rm B}^a$ & Electromagnetic field \\
% $G$ & Gauge group \\
$(\rho_I)^A_B$ & Gauge algebra generators \\
$f^I{}_{JK}$ & Structure constants\\
$\varepsilon^I$, $\varepsilon^\alpha$, $\varepsilon^{\alpha\beta}$ & Gauge group parameters \\
$J^i$, $\Sigma_\alpha{}^i$, $S_{\alpha\beta}{}^i$ & Noether currents \\
${\stackrel c J}{}^i$, ${\stackrel c \Sigma}{}_\alpha{}^i$, 
${\stackrel c S}{}_{\alpha\beta}{}^i$ & Convective currents \\
$M^{ij}$, $M_{\alpha}{}^{ij}$, $M_{\alpha\beta}{}^{ij}$ & Dipole moments \\
${\cal L}$, $L$, $L_D$,  $L_C$ & Lagrangian \\
$\gamma^\alpha$, $\beta$, $\alpha^a$, $\gamma_5$, $\sigma^{\alpha\beta}$ &  Dirac matrices \\
$\Psi$, $\psi$ & Dirac fermion field \\
\hline
\multicolumn{2}{l}{{Operators}}\\
\hline
${\cal H}$, ${\cal H}_{\rm FW}$ & Hamiltonian \\
$\epsilon$, $\epsilon'$ & Energy operator \\
$\bm r$ & Position operator \\
$\bm p$, $\bm\pi$ & Momentum operator \\
$\bm\lambda$ & Orbital moment\\
$\bm\Sigma$, $\bm\Pi$ & Spin, polarizaiton matrix\\
$\bm\Omega$, $\bm{\Omega}^{(T)}$ & Precession angular velocity\\ 
\hline
\multicolumn{2}{l}{{Auxiliary quantities}}\\
\hline
$\mu_0$, $\mu_N$, $\mu'$ & Magnetic moment \\
$\delta'$, $\rho'$, $\tau'$ & Coupling constants \\
$m$ & Fermion mass \\
$q$, $e$ & Electric charge \\
$\bm\omega$ & Angular velocity \\
$\gamma$ & Lorentz factor \\
$\nu_1$, $\nu_2$ & Zeeman frequencies \\
$g_N$, $g_1$, $g_2$ & $g$-factor \\
${\cal G}$ & Ratio of $g$-factors
\end{tabular}
\end{ruledtabular}
\end{table}
 
Our main conventions and notations are the same as in Refs. \cite{HonnefH,OST,OSTRONG,ostgrav}. In particular, the world indices are labeled by Latin letters $i,j,k,\ldots = 0,1,2,3$ (for example, the local spacetime coordinates $x^i$ and the holonomic coframe $dx^i$), whereas we reserve Greek letters from the beginning of the alphabet for tetrad indices, $\alpha,\beta,\ldots = 0,1,2,3$ (e.g., the anholonomic coframe $\vartheta^\alpha$). Furthermore, spatial indices are denoted by Latin letters from the beginning of the alphabet, $a,b,c,\ldots = 1,2,3$. In order to distinguish separate tetrad indices we put hats over them. 

We use the standard symbols $\wedge$ and $^\ast$ to denote the exterior product and the Hodge duality operator, respectively. The metric of the Minkowski spacetime reads $g_{\alpha\beta} = {\rm diag}(c^2, -1, -1, -1)$; the totally antisymmetric Levi-Civita tensor $\eta_{\alpha\beta\mu\nu}$ has the only nontrivial component $\eta_{\hat{0}\hat{1}\hat{2}\hat{3}} = c$. 

For Dirac matrices as well as for the gauge-theoretic notions and objects (including electrodynamics) we use the conventions of Bogoliubov-Shirkov \cite{BS}.

A directory of symbols used throughout the text can be found in Table \ref{tab_symbols}.

\end{document}